\documentclass{pas}

\usepackage{multirow}

\begin{document}

\lefttitle{Publications of the Astronomical Society of Australia}
\righttitle{Forbes, D. A. et al.}

\jnlPage{xx}{xx}
\jnlDoiYr{2024}
\doival{10.1017/pasa.xxxx.xx}

\articletitt{Research Paper}

\title{Keck/KCWI Spectroscopy of Globular Clusters in Local Volume Dwarf Galaxies}

\author{Duncan A. Forbes$^{1}$, 
Daniel Lyon$^2$, Jonah Gannon$^1$, Aaron J. Romanowsky$^{3,4}$ and Jean P. Brodie$^{1,2}$}

\affil{$^1$Centre for Astrophysics \& Supercomputing, 
Swinburne University, Hawthorn VIC 3122, Australia} 
\affil{$^2$School of Chemistry and Physics, Science and Engineering Faculty, Queensland University of Technology, Gardens Point Campus, Brisbane, QLD 4001, Australia}
\affil{$^{3}$Department of Physics and Astronomy, San Jos\'e State University, One Washington Square, San Jose, CA 95192, USA}
\affil{$^{4}$Department of Astronomy \& Astrophysics, university of California, Santa Cruz, CA 95064, USA}

\corresp{D. A. Forbes, Email: dforbes@swin.edu.au}

\citeauth{D. A. Forbes et al. Globular Clusters in Local Volume Dwarf Galaxies 
{\it Publications of the Astronomical Society of Australia} {\bf 00}, 1--12. https://doi.org/10.1017/pasa.xxxx.xx}

\history{(Received xx xx xxxx; revised xx xx xxxx; accepted xx xx xxxx)}

\begin{abstract}

A number of nearby dwarf galaxies have globular cluster (GC) candidates that require spectroscopic confirmation.
Here we present Keck telescope spectra for 15 known GCs and GC candidates that may be associated with a host dwarf galaxy, and an additional 3 GCs in the halo of M31 that are candidates for accretion from a now disrupted dwarf galaxy. We confirm 6 star clusters (of intermediate-to-old age) to be associated with NGC~247. The vast bulk of its GC system remains to be studied spectroscopically.  
We also confirm the GC candidates in F8D1 and DDO190, finding both to be young star clusters. 
The 3 M31 halo GCs all have radial velocities consistent with M31, are old and very metal-poor. Their ages and metallicities are consistent with accretion from a low mass satellite galaxy. 
Finally, three objects are found to be background galaxies -- two are projected near NGC~247 and one  (candidate GCC7) is near the IKN dwarf. The IKN dwarf thus has only 5 confirmed GCs but still a remarkable specific frequency of 124. 

\end{abstract}


\maketitle

\section{Introduction}

Globular clusters (GCs) are among the oldest structures in the Universe with a range of metallicities, compact sizes (half-light radii $\sim$ 3 pc) and a log-normal luminosity function (Brodie \& Strader 2006). 
There are many outstanding issues concerning their formation and destruction, enrichment history and 
relationship with host galaxy properties (Forbes et al. 2018). They reveal a remarkable linear scaling relation between their number and the total halo mass of their host galaxy (e.g. Burkert \& Forbes 2020). For low GC numbers (in low mass galaxies) the relation has considerable scatter as stochastic effects dominate.

Less massive galaxies tend to host fewer GCs than more massive ones. 
However, one could argue that the importance of GCs increases for dwarf galaxies since 
the number of GCs per unit galaxy starlight (specific frequency, S$_N$), and the GC system mass to stellar mass ratio, can be much higher than for more luminous/more massive galaxies (e.g. Georgiev et al. 2010; Forbes et al. 2020). Once formed in a vigorous star formation event, GCs are imprinted with a formation age and metallicity that is unchanged over cosmic time. 
The GCs of dwarf galaxies provide important assembly history clues of not only their host galaxy (Georgiev et al. 2010), but also giant galaxies since 
their metal-poor GC subpopulation is thought to be largely accreted from disrupted dwarf galaxies (Forbes \& Remus 2018).


The low number of GCs in dwarfs means that it is very important to confirm the status of each given the 
addition/removal of a single GC candidate can make a large difference to its GC system (in some cases the system is either one GC or none). Although imaging searches have been carried out for many local dwarfs (e.g. Georgiev et al. 2010), spectroscopy is needed to confirm a common radial velocity with the host galaxy and an old age (i.e. $\ge$ 8 Gyr) to confirm its status as a GC rather than young massive cluster. 


Follow-up spectroscopy is time consuming and requires accurate coordinates from the imaging study. We note here the incorrect coordinates for a single GC candidate in the nearby dwarf galaxy KK65, imaged by HST, as reported by Sharina et al. (2005). Their coordinates for KK65-3-1095 are 
07:42:29.4, +16:34:29 (J2000). Examining WFPC2 and ACS imaging in the Hubble Legacy Archive reveals that no bright object lies at these coordinates. Sharina et al. (2005) indicate that the GC is located within the main body of the galaxy, projected 0.33 kpc from its centre. There is indeed a single resolved GC-like object within the body of the galaxy. It is however located at 07:42:31, +16:33:30 (J2000), and deserves spectroscopic follow-up.

Here we present spectroscopy of some known GCs and additional GC candidates using the Keck Cosmic Web Imager (KCWI) integral field unit on the Keck II telescope. The targets were selected to be visible during twilight and/or during intervals in our main observing programmes. We focused on blue GCs which might be metal-poor. The current record holder for a metal-poor GC is EXT8 in the halo of M31 with [Fe/H] = --2.90 (Larsen et al. 2020). Prior to this measurement it was thought that a floor exists in metallicity of [Fe/H] $\sim$ --2.5 for GCs. 
Our targets include GC candidates in NGC~247 (D = 3.3 Mpc), DDO190 (D = 2.9 Mpc), IKN (D = 3.8 Mpc), along with known GCs in F8D1 (D = 3.8 Mpc) and Sextans~A (D = 1.4 Mpc). We also include 3 known GCs in the halo of M31 (which may have been accreted from a dwarf satellite galaxy). As well as deriving radial velocities (to confirm association with the possible host galaxy) we derive stellar population properties and compare the GCs in an age-metallicity diagram.

\section{Data}

\subsection{The Targets}

In Figure 1 we show a montage of the 15 target objects. The imaging comes from HST/ACS, or, if not available we show  colour images from Legacy (https://www.legacysurvey.org/dr9/description/) or SDSS 
(https://classic.sdss.org/home.php) surveys.
We have 8 targets for NGC~247: 6 with HST imaging and 2 with only ground-based Legacy imaging. HST imaging exists for the sources in DDO190 and  F8D1, and for two in M31 (PA-41 and B336). In the latter two cases, the objects are located in the halo of M31 and clearly resolved with individual bright stars seen in their outer regions. For H12 in the halo of M31 and the known GC around Sextans~A we only have ground-based imaging. Further details of the individual targets, and their possible host galaxy, are discussed below.

\subsection{Data Acquisition}

We observed 15 known GCs and GC candidates using the Keck Cosmic Web Imager (KCWI) on the Keck II telescope over several years. 
Table 1 summarises the observations. It lists the target object, its possible host galaxy and its B-band absolute magnitude and stellar mass (with dwarf galaxy values from Weisz et al. 2011), coordinates of the target, date of observation, KCWI slicer used (M = Medium, L = Large), Grating (BL, BH3 or BM), central wavelength setting and exposure time. 

\subsection{Data Reduction}

All data were reduced using the standard KCWI data reduction pipeline (Morrissey et al. 2018), following the method described in Gannon et al. (2020). After reduction, data cubes were cleaned and trimmed to their good spatial and spectral wavelength ranges and then processed as below. 
The output was standard star calibrated, non-sky subtracted data cubes. 
The reduced KCWI datacubes were displayed using Qfitsview. A aperture was defined to include the total light from each GC. 
A background annulus (which includes both sky and galaxy backgrounds) was also defined around each GC and subtracted. A 1D spectrum was then extracted. 
Barycentric corrections were then applied and each spectrum was median combined to produce the final science spectrum.
The final spectra have S/N ratios $\sim$20 per \AA, sufficient for deriving both radial velocities and stellar populations. The exceptions are the spectra of GC candidates in 
DDO190 and F8D1 which have a lower S/N of $\sim$10 per \AA. 


\subsection{Data Analysis}

We used the code \texttt{pPXF} (Cappellari  et al. 2017) to measure recessional velocities and stellar populations from each spectrum. We follow the method described in Gannon et al. (2020) which involves fitting the spectrum using the Coelho et al. (2014) synthetic stellar library in 256 different input parameter combinations to \texttt{pPXF}. A modest range of additive and multiplicative polynomials were used. When present, emission lines are 
simultaneously modelled with the absorption features.


\begin{table*}[b!]
 \caption{Observations. Host properties and GC candidates coordinates and KCWI setup.}\label{sample-table}
 {\tablefont\begin{tabular}{@{\extracolsep{\fill}}lccccccccccc}
   \toprule
    Target & Host Galaxy & Host M$_B$ & Host M$_{\ast}$ & R.A. & Dec. & Date & Slicer & Grating & R & Central Wavelength & Exp.Time \\
    & & (mag) & (10$^8$ M$_{\odot}$) & (J2000) & (J2000) & & & & &(\AA) & (s) \\
     \hline
GC1 & Sextans~A & -13.7 & 1.4 & 10:10:43.8 & -04:43:28.9 & 2020/11/10 & M & BM & 4000 & 4925 & 2420\\
GC1 & NGC~247 & -19.2 & 30 & 00:47:07.6	& -20:45:29.1 & 2023/10/23 & M & BL & 1800 & 4550 & 6240\\
GC2 & NGC~247 & -19.2 & 30 & 00:47:07.6	& -20:45:20.3 & 2023/10/23 & M &BL & 1800 & 4550 & 6240\\
GC3 & NGC~247 & -19.2 & 30 & 00:47:07.5 & -20:45:19.8 & 2023/10/23 & M &BL & 1800 & 4550 & 6240\\
GC4 & NGC~247 & -19.2 & 30 & 00:47:13.4 & -20:47:22.8  & 2023/10/23 & M &BL & 1800 & 4550 & 5040\\
GC5 & NGC~247 & -19.2 & 30 & 00:47:14.2 & -20:47:22.1 & 2023/10/23 & M &BL & 1800 & 4550 & 5040\\
GC6 & NGC~247 & -19.2 & 30 & 00:47:14.2 & -20:47:17.7 & 2023/10/23 & M &BL & 1800 & 4550 & 5040\\
5P & NGC~247 & -19.2 & 30 & 00:46:54.6 & -20:41:29.5 & 2023/10/23/ & M &BL & 1800 & 4550 & 2520 \\
10P & NGC~247 & -19.2 & 30 & 00:47:09.9 & -20:30:19.0 & 2023/10/23 & M &BL & 1800 & 4550 & 2520\\
GCC1 & DDO190 & -14.1 & 0.6 & 14:24:45.0 & 44:31:36.1 & 2020/02/16 & M &BH3 & 9000 & 5080 & 1200\\
GC1 & F8D1 & -12.2 & 1.4 & 09:44:39.4 & 67:26:05.9 & 2021/01/13 & M &BL & 1800 & 4550 & 1800 \\
GCC7 & IKN & -10.8 & 2.0 & 10:08:08.9 & 68:28:36.8 & 2021/02/17 & L &BL & 900 & 4550 & 3600\\
B336 & M31 & -21.2 & 1000 & 00:40:47.6 & 42:08:41.8 & 2020/11/10 & M &BH3 & 9000 & 5070 & 3600\\
H12 & M31 & -21.2 & 1000 & 00:38:03.6 & 37:43:59.0 & 2020/10/20 & M & BH3 & 9000 & 5070 & 5400 \\
PA-41 & M31 & -21.2 & 1000 & 00:53:39.6 & 42:35:15.0 & 2022/01/29 & L &BL & 900 & 4550 & 900\\
   \botrule
    \end{tabular}}
    \begin{tabnote}
    \end{tabnote}
\end{table*}

\begin{figure*}[t]
  \centerline{\vbox to 6pc{\hbox to 10pc{}}}
  \includegraphics[scale=0.15]{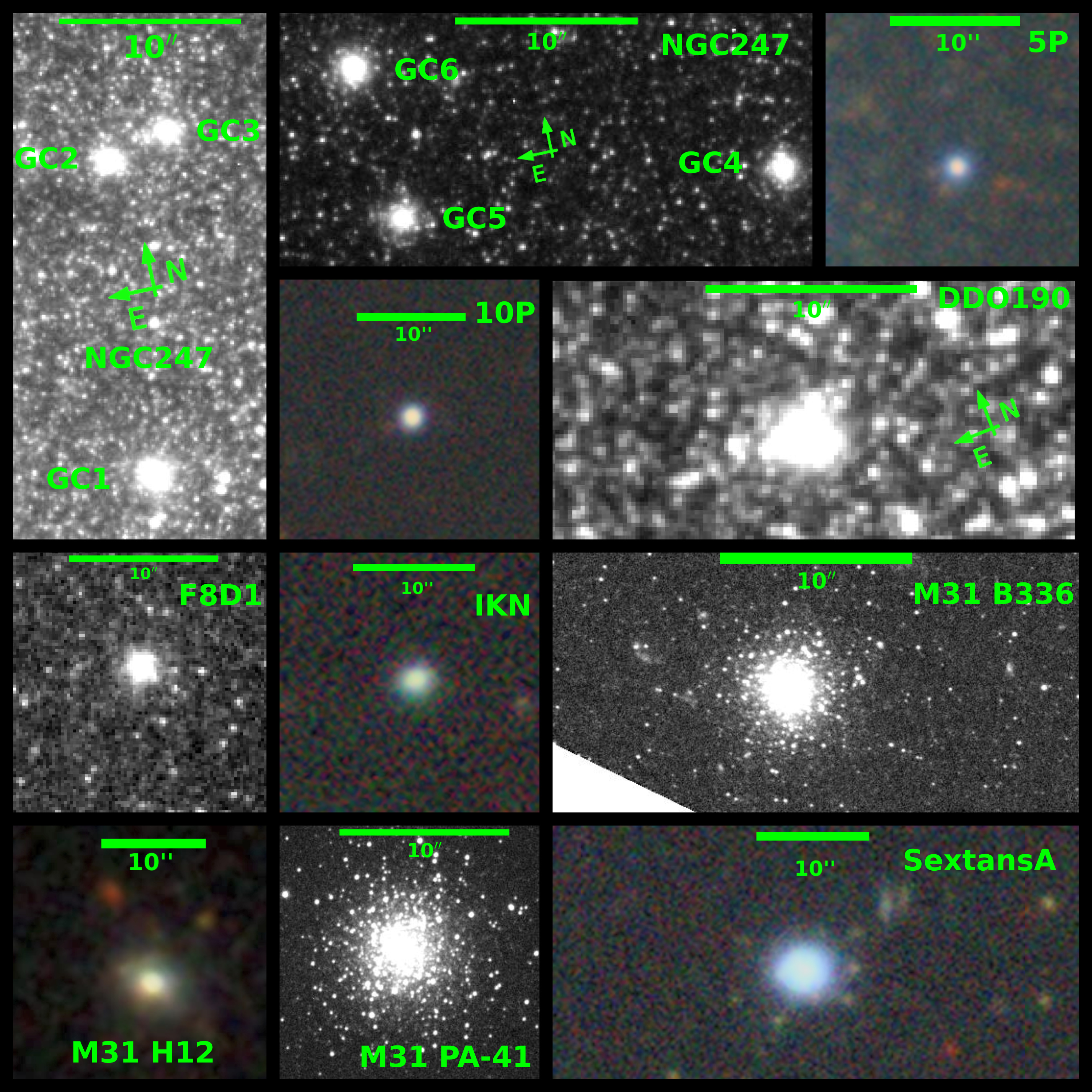}
  \caption{Montage of sample GCs and GC candidates. From left to right and down are images of NGC 247, DDO190, F8D1, IKN, M31 and Sextans~A sources. Single filter HST/ACS imaging is shown if available, otherwise colour images are from Legacy or SDSS ground-based imaging. The orientation of the images is North up, East left unless otherwise shown. A 10'' bar is shown in each panel.}
  \label{sample-figure}
\end{figure*}

\section{Results and Discussion}

In Figures 2 and 3 we present spectra of the star clusters (i.e. old GCs and younger star clusters) associated with their host galaxy.
As well as the original spectrum (black) we show the ppxf best fit model (magenta). 
Figure 4 shows the 3 GC candidates that we find to be background galaxies. These candidates are projected on the sky near NGC~247 (5P and 10P) and the IKN galaxy.  
The wavelength range and resolution of the spectra varies (see Table 1 for setup used) but each spectrum always includes H$\beta$. 

\begin{figure}[t]
  \centerline{\vbox to 6pc{\hbox to 10pc{}}}
  \includegraphics[scale=0.5]{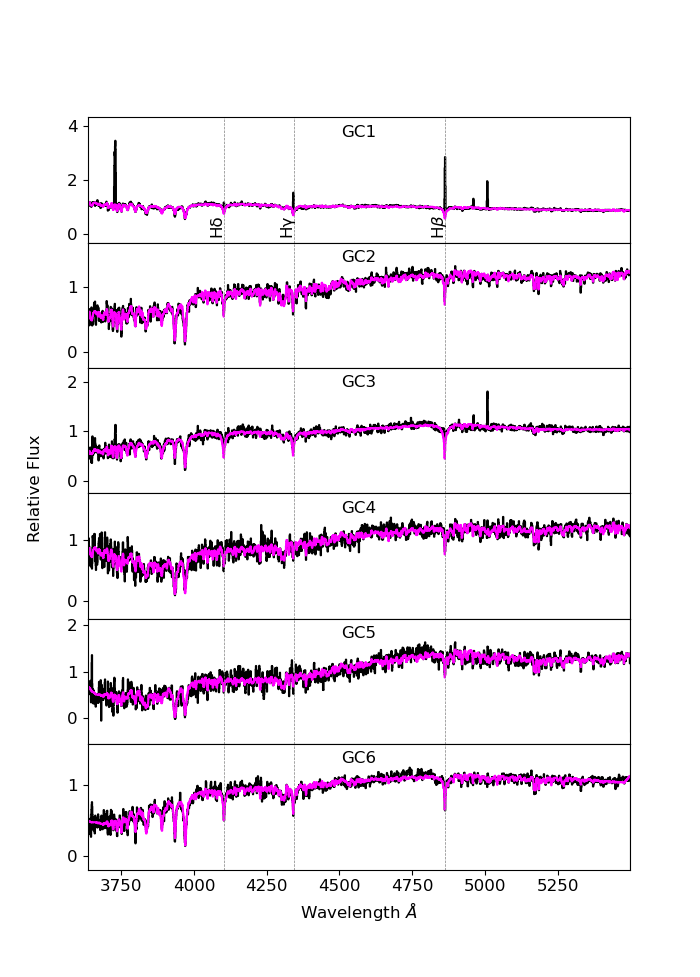}
  \caption{KCWI spectra of 6 GC candidates around NGC~247. They range from young massive star clusters to old globular clusters, all associated with NGC~247. Emission lines in the spectra of GC1 and GC3 may be the result of incomplete galaxy background subtraction. Key absorption lines are highlighted. }
  \label{sample-figure}
\end{figure}

\begin{figure}[t]
  \centerline{\vbox to 6pc{\hbox to 10pc{}}}
  \includegraphics[scale=0.5]{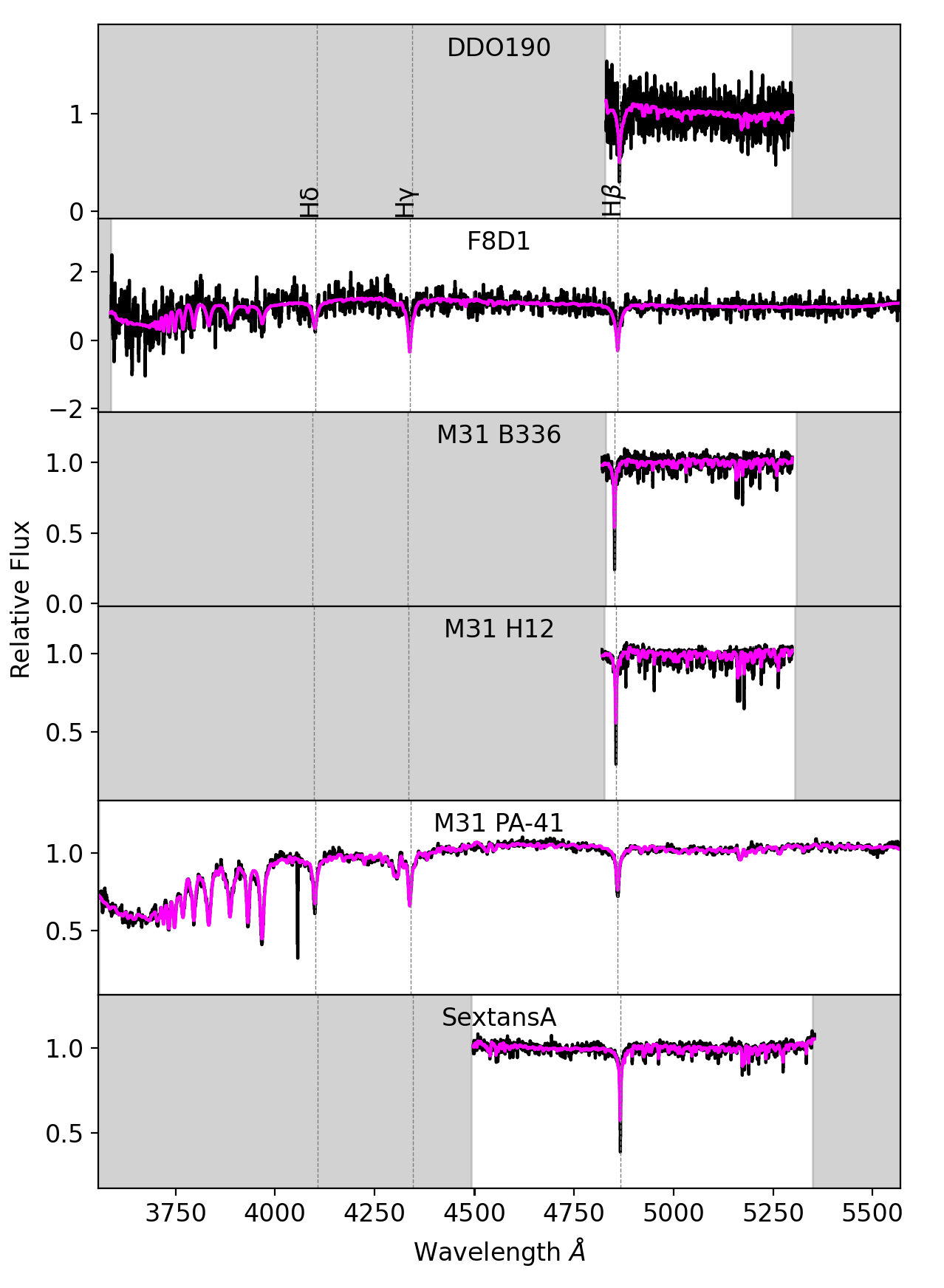}
  \caption{KCWI spectra of GCs associated with other galaxies. They include GCs associated with DDO190, F8D1, M31 and Sextans~A. Key absorption lines are highlighted. }
  \label{sample-figure}
\end{figure}

\begin{figure}[t]
  \centerline{\vbox to 6pc{\hbox to 10pc{}}}
  \includegraphics[scale=0.5]{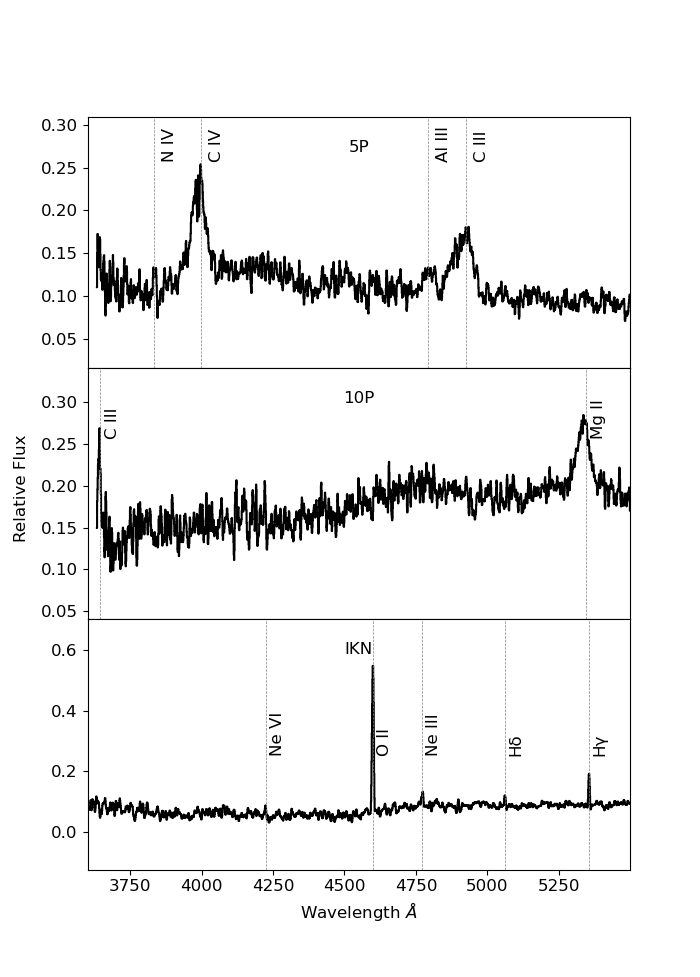}
  \caption{KCWI spectra of three GC candidates that are background galaxies. Candidates are 5P and 10P projected near NGC~247 and GCC7 near the IKN galaxy. Key emission lines are highlighted. }
  \label{sample-figure}
\end{figure}

In Table 2 we summarise our results. This includes the radial velocity (RV) and stellar population parameters, along with uncertainties. 
The radial velocity errors we report only include the formal measurement uncertainty from ppxf and do not include systematics which would be on the order of 10--30 km/s. Stellar population results come from the median of the parameter distributions with uncertainties taken as the 16th and 84th percentiles of these distributions. 
Below we discuss our results for individual galaxies. 
We note the caveat that any integrated light spectra of metal-poor GCs can be affected by the presence of horizontal branch stars,  
which may bias results towards to younger ages.\\

\begin{table}[b!]
 \caption{Results. Recession Velocities and Stellar Populations (and errors). }\label{sample-table}
 {\tablefont\begin{tabular}{@{\extracolsep{\fill}}lccccc}
   \toprule
    Target & Host & RV & Age & [Z/H] & Comment\\
    & & (km/s) & (Gyr) & (dex) & \\
     \hline
GC1 & Sextans~A & 296 (2.0) & 8.3 (+1.2, -0.9) & -2.11 (+0.06, -0.03) & \\
GC1 & NGC~247 & 83 (4.6) & 5.0 (+0.4,-0.7) & -1.27 (+0.25, -0.15) & \\
GC2 & NGC~247 & 54 (2.8) & 7.7  (+0.6, -0.7) & -1.03 (+0.09, -0.05) & \\
GC3 & NGC~247 & 41 (6.0) & 4.3 (+0.4, -0.5) & -2.04 (+0.22, -0.07) & \\
GC4 & NGC~247 & 77 (5.1) & 9.2 (+0.5, -0.7) & -0.79 (+0.06, -0.05) & \\
GC5 & NGC~247 & 86 (5.0) & 10.6 (+0.7, -2.0) & -0.49 (+0.07, -0.19) &\\
GC6 & NGC~247 & 91 (4.0) & 6.7 (+0.4, -1.0) & -1.59 (+0.08, -0.04) & \\
5P & NGC~247 & z = 1.58 & -- & -- & AGN\\
10P & NGC~247 & z = 0.91 & -- & -- & AGN\\
GCC1 & DDO190 & 160 (7.0) & 2.6 (+1.8, -1.3) & -2.17 (0.12, -0.05)  &\\
GC1 & F8D1 & -108 (23) & 0.48 (+0.5, -0.2) & -1.06 (+0.56, -0.55) & \\
GCC7 & IKN & z = 0.23 & -- & -- & galaxy\\
B336 & M31 & -595 (2.3) & 7.0 (+0.6, -0.7) & -2.01 (+0.06, -0.05) & \\
H12 & M31 & -379 (1.9) & 8.2 (+0.8, -1.0) & -1.88 (+0.05, -0.04)  & \\
PA-41 & M31 & -104 (4.4) & 7.8 (+1.5, -1.3) & -1.93 (+0.08, -0.06) & \\
\botrule
    \end{tabular}}
    \begin{tabnote}
    \end{tabnote}
\end{table}

\subsection{Individual Galaxies}

\noindent
$\bullet$ {\bf Sextans~A}\\
Sextans~A is classified as IBm with a radial velocity of 324 $\pm$ 1 km/s from NED. Its single GC has been studied by Beasley et al. (2019). They found RV = 305 $\pm$ 15 km/s, age = 8.6 $\pm$ 2.7 Gyr and [Fe/H] = --2.38 $\pm$ 0.29. The more recent work by (Gvozdenko et al. 2024) found the same age with a metallicity [Fe/H] = --2.14 $\pm$ 0.04.  Thus it is an old, very metal-poor GC associated with this dwarf galaxy. From our KCWI spectrum we measure RV = 296 $\pm$ 2.0 km/s and 
stellar populations that are similar to those reported, i.e. age = 8.3$^{+1.2}_{-0.9}$ and [Z/H] = --2.11$^{+0.06}_{-0.03}$. This gives us further confidence in our stellar population fitting. 
We note that the total metallicity is expected to be similar, or higher by $\sim$0.3 dex, than the iron metallicity depending on the (unknown) alpha-element abundance. \\

\noindent
$\bullet$ {\bf NGC~247}\\
NGC~247 is an edge-on SBd spiral in the nearby Sculptor Group, with ongoing star formation activity and a stellar mass of $\sim 3\times 10^9 M_{\odot}$. 
From a spectrum of the nucleus and using ppxf, Kacharov et al. (2018) determined its star formation history and found it to be very broad. At old ages (7--10 Gyr) it displays a wide range of metallicities (--2.3 $<$ [Fe/H] $<$ --0.5). This enriches to --1 $<$ [Fe/H] $<$ --0.5 around 2--3 Gyr ago. However, there are also some indications of star formation $\sim$ 10 Gyr ago that was metal-rich ([Fe/H] $\sim$ 0), and $\le$ 1 Gyr ago that was metal-poor ([Fe/H] $\sim$ --1.5). Thus it appears to have quite a complex star formation/chemical enrichment history.

The galaxy has a radial velocity of 156 $\pm$ 2 km/s. To our knowledge there are only 3 spectroscopically confirmed GCs from Olsen et al. (2004). Their GCs, named NGC247-1, -2 and -3 (in their table 4) have metallicities of [Fe/H] = --1.13, --0.41 and --0.89 respectively. No ages were reported. 
They estimated a total of 60 $\pm$ 30 GCs, so that the majority remain to be confirmed. 
A young `fuzzy' star cluster with an age of $\sim$ 300 Myr and [Z/H] $\sim$ --0.6 was reported recently by Romanowsky et al. (2023). 

Using imaging from the Subaru telescope, Santhanakrishnan (2016) identified 14 additional GC candidates (named 1P -- 14P). An additional 6 candidates were initially identified in the Subaru imaging and these were found to be included in a single HST/ACS image of NGC 247 (PI: Dalcanton). These  objects appear in the HST imaging as star clusters with resolved stars in their outer regions. See Romanowsky et al. (2023) for further details of the Subaru imaging. 
Using KCWI we obtained data for all 6 HST star clusters (in two pointings) and for the candidates 5P and 10P (in two separate pointings), giving a total of 8 GC candidates. 
Of the HST sources, we find all 6 to be bona fide star clusters associated with NGC~247. They have a range of ages and metallicities 
4.3 $<$ age $<$ 10.6 Gyr
--0.49 $<$ [Z/H] $<$ --2.0. Several of them might be better described as intermediate-aged  star clusters. We note that GC1 and GC3 have emission lines from imperfect galaxy background subtraction, however these were excluded in the ppxf fit. 
Two candidates (those from the ground-based imaging), 5P and 10P, are found be to background AGN and not GCs. 

A relevant comparison galaxy is NGC 2403 which lies at a distance of 3.2 Mpc and has a stellar mass of 7 $\times$ 10$^9$ M$_{\odot}$. Using KCWI spectroscopy and a similar analysis process we measured a range of ages (0.4 $<$ age $<$ 12.5 Gyr) and metallicities (--0.61 $>$ [Z/H] $>$ --2.07) for half a dozen star clusters (Forbes et al. 2022). Thus similarly to NGC 247, the star clusters of NGC 2403 appear to have formed in-situ over an extended period of time.   
\\

\noindent
$\bullet$ {\bf DDO190}\\
DDO190 (UGC9240) is an Im dwarf with a RV = 150 $\pm$ 4 km/s. Sharina et al. (2005) identified 1 very blue candidate in HST/WFPC2 imaging which they named U9240-3-4557 (we use GCC1 here). They resolved the candidate and measured  a half-light radius of 3.6 pc and M$_V$ = --7.2. 
From our KCWI spectrum of this candidate we measure RV = 160 $\pm$ 7 km/s. This RV is well within the velocity range of the rotating HI gas (W$_{50}$ = 45 km/s) in DDO190 (Ott et al. 2012) and thus we confirm a GC associated with the galaxy. Our spectrum (Figure 3) is limited in wavelength and of low S/N but an young age (2.6 Gyr) and low metallicity ([Z/H] = --2.17) is indicated (as might be expected from its blue colour).\\

\noindent
$\bullet$ {\bf F8D1}\\
F8D1 is a low surface brightness dwarf galaxy in the M81 group (D = 3.8 Mpc). The galaxy has been studied using HST imaging by Zemaitis et al. (2023). It reveals a tidal tail indicating an ongoing interaction and a TRGB distance of 3.67 Mpc. It has a metallicity [M/H] = --1.14 $\pm$ 0.09. Its large half-light radius and low surface brightness make it an ultra-diffuse galaxy (UDG). 
F8D1 has a single GC has a spectroscopic RV measured by Chiboucas et al. (2009) of --125 $\pm$ 130 km/s. The RV for the galaxy is reported in NED as --125 with a reference to 
Kourkchi \& Tully  (2017). We suspect the latter is actually the RV of the GC as reported by Chiboucas et al. (2009) and not the host galaxy. The M81 group has a range of RVs with M81 itself at RV = --36 $\pm$ 3 km/s.  We measure --108 $\pm$ 23 km/s for the GC, which we associate with F8D1. Our low S/N spectrum of F8D1 GC reveals several strong Balmer lines and we find a very young age of 480 Myr. Its recent formation suggests that it has formed as result of the ongoing interaction.\\

\noindent
$\bullet$ {\bf IKN}\\
The IKN dwarf, in the M81 group (D = 3.8 Mpc),  is a very low surface brightness dwarf and a candidate for a UDG. It has 5 confirmed 
old GCs (Tudorica et al. 2015) giving it a remarkable 
specific frequency of S$_N$ = 124 (Georgiev et al. 2010).
Based on SDSS DR10 imaging, Tudorica et al. (2015) identified 2 GC candidates requiring spectroscopic follow-up. Candidate GCC6 lies very close to a bright star. Here we present a spectrum of GCC7 and find it to be a background galaxy at z $\sim$ 0.2. \\

\noindent
$\bullet$ {\bf M31}\\
Three known GCs in M31 were observed. They are B336, H12 and the outer halo GC PA-41 from the PAndAS survey (Huxor et al. 2014). 
The Revised Bologna Catalog (RBC) version 5
http://www.bo.astro.it/M31/) lists B336 with a RV = --609 $\pm$ 27 km/s. Its metallicity is very low but with a large error, i.e.
[Fe/H] = --2.5 $\pm$ 0.6. 
For H12, Zhou et al. (2011) measured RV = --412 $\pm$ 33 km/s and 
old ages (13.6$^{+1.4}_{-1.7}$ Gyr), metal-poor ([Fe/H] = --1.80$\pm$0.18) and alpha enhanced ([$\alpha$/Fe] $\sim$ 0.5) in their Cas system. 
PA--41, also known as SDSS9, has a very blue colour of (g--i)$_0$ = 0.69 from de Tullio Zinn \& Zinn (2013), suggesting it might also be quite metal-poor. 
We are not aware of any recession velocity reported for PA-41. According to Mackey et al. (2019), PA-41 is likely associated with a stellar stream (C/D) that came from a disrupted satellite galaxy. 
We confirm association with M31 for all 3 GCs and 
derive old ages of $\sim$8 Gyr and very low metallicities [Z/H] $\sim$ --2 (although not as low as EXT8). 

\subsection{Age-metallicity Relation}

The ages and metallicities of GCs formed within a galaxy broadly follow an age-metallicity relation (AMR) that describes the chemical enrichment in that galaxy over cosmic time (e.g. Horta et al. 2021), although significant variations due to local conditions can be present (e.g. infalling halo gas may lower the mean metallicity of in-situ gas).  
An AMR can be approximated by a closed box chemical enrichment model. 
For example, Forbes (2020) assigned Milky Way GCs to several disrupted dwarf satellite galaxies on the basis of their ages and metallicities. The corresponding AMRs could then be used to estimate the original mass of the disrupted dwarf.

\begin{figure}[t]
  \centerline{\vbox to 6pc{\hbox to 10pc{}}}
  \includegraphics[scale=0.33, angle=-90]{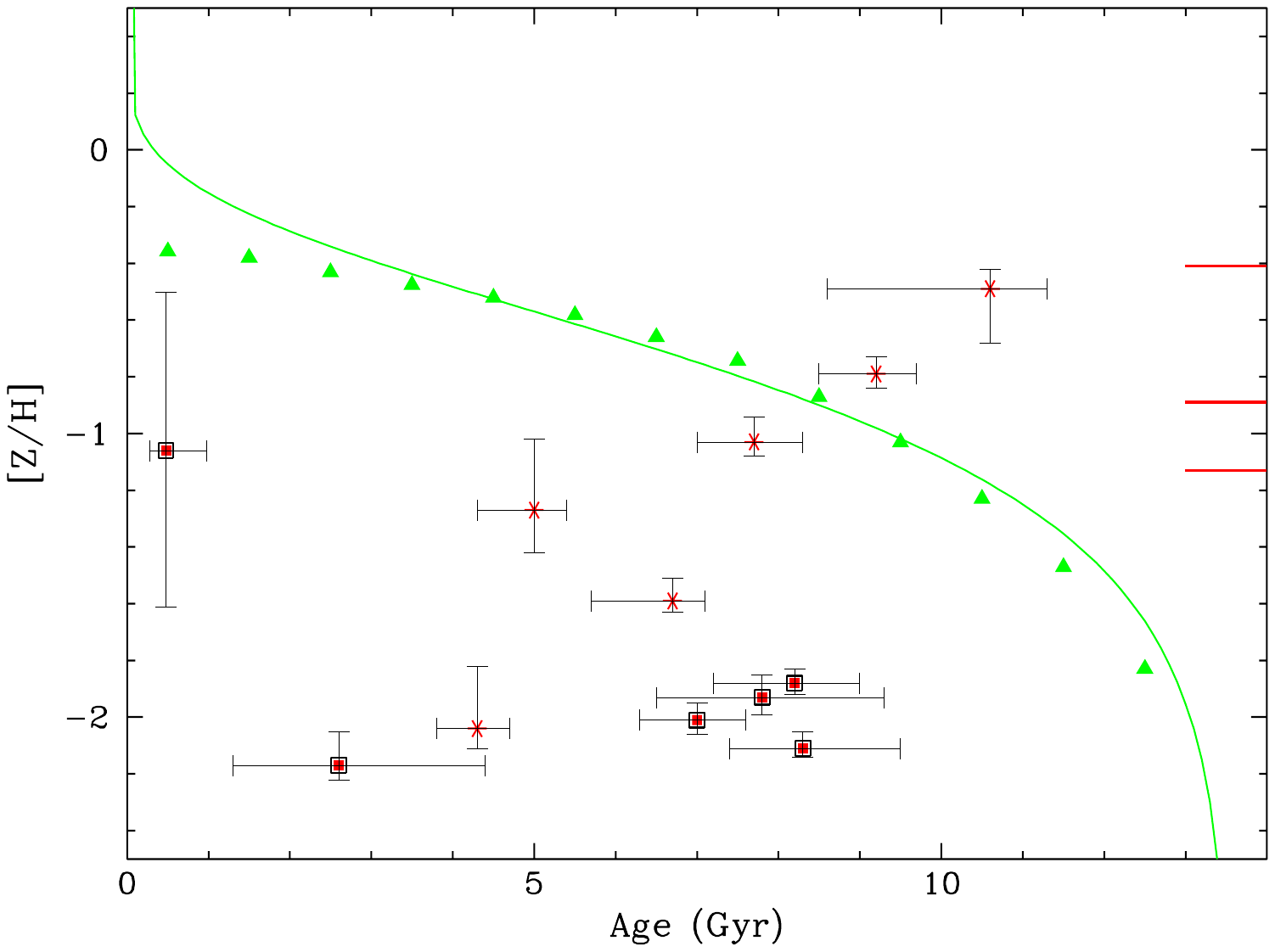}
  \caption{Age-metallicity diagram of sample GCs. red asterisks show GCs associated with NGC~247, and red squares with other host galaxies. The 3 red lines on the right hand side indicate the metallicities of 3 GCs in NGC 247 from Olsen et al. (2004). The solid line shows the age-metallicity relation inferred for Gaia-Enceladus dwarf galaxy from Forbes (2020) and triangles for the  model AMR from Horta et al. (2021) for a stellar mass $\sim10^9$ M$_{\odot}$ galaxy. The NGC~247 star clusters show a broad range of age and metallicity similar to that reported for the nuclear region by Kacharov et al. (2018). The GCs associated with other galaxies tend to be less enriched than the AMRs shown, consistent with 
  formation in a low mass ($<$ 10$^9$ 
  M$_{\odot}$) galaxy.}
  
  \label{sample-figure}
\end{figure}
In Figure 5 we show our age and metallicity measurements for the old GCs and younger star clusters. We also show the AMR of the Gaia-Enceladus dwarf galaxy (we have assumed that [Z/H] $\sim$ [Fe/H]). This dwarf is thought to have had a  stellar mass of $\sim10^9$ M$_{\odot}$, hosted $\sim28$ GCs and was accreted by the Milky Way some 10 Gyr ago (Forbes 2020). We also include the AMR model by Horta et al. (2021) for a galaxy with a similar stellar mass. 
NGC~247 has a 
stellar mass of 3 $\times$ 10$^9$ M$_{\odot}$. The plot shows 
that the NGC~247 star clusters reveal a broad range of ages and metallicities similar to that reported by Kacharov et al. (2018). 
We also include markers for the 3 NGC~247 GCs with [Fe/H] metallicities (ages were not published) from Olsen et al. (2004). Given that the total GC system may contain up to 60 $\pm$ 30 GCs (Olsen et al. 2004), it would be worthwhile to obtain more spectroscopic 
samples to investigate whether distinct age-metallicity relations exist for old GCs vs the more recently formed star clusters. 

The GCs associated with other galaxies have a wide range in ages and metallicities. They all occupy a location in the diagram below the AMR of a 10$^9$ M$_{\odot}$ galaxy. For the 3 M31 halo GCs, this suggests that if they formed in a now accreted dwarf satellite galaxy that this galaxy had an original mass $<$ 10$^9$ M$_{\odot}$. 

\section{Summary}

Here we present KCWI spectra of 15 known GCs and GC candidates in several nearby dwarf galaxies and in the halo of M31. We derive radial velocities and ages and metallicities for each object.

For the known GC associated with Sextans~A we find a  similar radial velocity and stellar populations to Beasley et al. (2019), of an old age and very low metallicity. 
Of the 8 GC candidates that might be associated with NGC~247, we find 2 (5P and 10P) to be background AGN.  The other 6 candidates around NGC~247 are confirmed as star clusters with a range from ages from intermediate to old (i.e. bona fide GCs). They also reveal a range of metallicities and scatter widely about an age-metallicity relation.
For the IKN galaxy the 1 GC candidate is found to be a z $\sim$ 0.2  background galaxy. Thus IKN's total GC system remains at 5, which still has a remarkable specific frequency of S$_N$ = 124. For the GC candidate in DDO190 confirm its association. Our low S/N spectrum indicates a young age ($\sim$ 3 Gyr) and low metallicity ([Z/H] $\sim$ --1). This suggests that the candidate is a young star cluster (consistent with its very blue colour). 
For the known GC in F8D1, we measure a more accurate velocity confirming association. Its stellar population properties, again from a low S/N spectrum, suggests very recent formation as a result of the ongoing tidal interaction. 
The 3 M31 halo GCs are confirmed by their radial velocities and are found be to very metal-poor and old. They may have been accreted from a disrupted low mass ($<$ 10$^9$ M$_{\odot}$) dwarf galaxy. 


\subsection{Bibliography}

\section*{Data Availability}

Raw data is available in the Keck Observatory Archive (KAO). 

\section{Acknowedgements}

We thank the referee for several useful comments. 
We thank L. Buzzo, L. Haacke, S. Janssens, A. Ferre-Mateu for help and useful discussions. We thank J. Carlin for initial ground-based images of NGC 247 that aided in the selection of spectroscopic targets. We thank V. Santhanakrishnan for 
identifying GC candidates around NGC~247, and S. Larsen and 
A. Gvozdenko for M31 targets. 
This work was supported by a NASA Keck PI Data Award, administered by the NASA Exoplanet Science Institute. We thank the ARC for financial support via DP220101863. 
Here we make use of NED (https://ned.ipac.caltech.edu/) and Qfitsview (https://www.mpe.mpg.de/ott/QFitsView/).  The data presented herein were obtained at Keck Observatory, which is a private 501(c)3 non-profit organization operated as a scientific partnership among the California Institute of Technology, the University of California, and the National Aeronautics and Space Administration. The Observatory was made possible by the generous financial support of the W. M. Keck Foundation. The authors wish to recognize and acknowledge the very significant cultural role and reverence that the summit of Maunakea has always had within the Native Hawaiian community. We are most fortunate to have the opportunity to conduct observations from this mountain.

\end{document}